\documentclass[12pt]{article}

\usepackage{hyperref}
\usepackage{amsmath}
\usepackage{amssymb}
\usepackage{graphicx}
\usepackage{color}
\usepackage{psfrag}
\usepackage{subfigure}
\begin{document}
\flushright{USTC-ICTS-11-11}
\begin{center}{\Large \bf
Bubbles and Black Branes in Grand Canonical Ensemble}
\\[2cm]
{\large Chao Wu\footnote{Email:wuchao86@mail.ustc.edu.cn}, Zhiguang
Xiao\footnote{Email:xiaozg@ustc.edu.cn}, and Jianfei
Xu\footnote{Email:jfxu06@mail.ustc.edu.cn}
\\[4mm]
{\em Interdisciplinary Center for Theoretical Study\\
 University of Science and Technology of China, Hefei, Anhui
 230026, China}
}
\end{center}
\vspace{2cm}
\abstract{
When the phase structure of the black brane in grand canonical
ensemble is discussed, the bubble phase with the same boundary data
should be included in this structure. As such, the phase transitions
among bubbles, black branes and ``hot flat space'' are possible,
therefore giving a much enriched phase structure.  We also argue that under
some conditions, either the grand canonical ensemble itself is
unstable or there are some unknown new phases.
}

\section{Introduction}
Understanding the thermodynamic phase structure of black holes is
helpful in learning the properties of space-time and quantum gravity.  
The phase structure of the AdS black holes can also be used to study
the corresponding field theory within the  context of AdS/CFT
correspondence. The well-known Hawking-Page phase transition
\cite{Hawking:1982dh} between the AdS black hole and the ``hot empty AdS
space'' corresponds to the confinement-deconfinement phase transition
of the large-$N_c$ ${\cal N}=4$ super Yang-Mills at finite temperature
\cite{Witten:1998zw}. The charged AdS black hole in the canonical
ensemble also displays a van der Waals--Maxwell-like phase transition which
can also be understood from the dual-field
theory\cite{Chamblin:1999tk,Chamblin:1999hg}. 

The Hawking-Page or van der Waals--Maxwell-like phase structure is not
only present in AdS black holes but also exists in some other black
holes. The Hawking--Page-like transition can also be found in the
canonical ensemble between chargeless asymptotic flat black holes and
``hot flat space'' as discussed by York \cite{York:1986it}. In the
grand canonical ensemble, there can also be transitions between 
charged black holes and the hot flat space \cite{Braden:1990hw} .
The van der Waals like phase transition was also seen in asymptotic
flat and dS black hole \cite{Carlip:2003ne,Lundgren:2006kt}. Unlike
the black holes in AdS space, asymptotic flat and dS black holes are
not thermodynamically stable objects due to Hawking radiation. So in
discussing the phase structure of these objects, the black hole must
be put in thermal contact with a heat reservoir with fixed thermodynamic data on the
boundary so that they can form a equilibrium, and then the phase
structure could be discussed.
 
Black branes are the black hole-like solutions of low-energy effective
theory of string or M-theory, which are also asymptotic flat. They
also need to be put inside a heat reservoir in order to study their
thermodynamic properties. In fact, they have similar phase transitions
as those asymptotic flat black holes.
In \cite{Lu:2010xt}, the phase structure of black $p$-branes in
$D$-dimension ($D=d+\tilde d +2, p=d-1$) space-time in canonical
ensemble is discussed, where the charge $q$ inside the cavity is fixed.
In this ensemble, for $\tilde d>2$ charged cases, there is a critical
charge $q_c$. When $q<q_c$, there is a first-order van der Waals
phase transition between the larger black brane solution and the
smaller one at a certain transition temperature.  However, the chargeless case
and $\tilde d\leq2$ charged cases are different from $\tilde d>2$
charged cases.  There is no such van der Waals phase transition and
below a minimal temperature $T_{\rm min}$, there could not exist a black brane
phase. For the chargeless case the hot flat space is the stable
phase below $T_{\rm min}$, whereas in the charged cases we do not know
what could exist at the present stage, because of the absence of the charged
hot flat space.  
One would wonder what could fill this part of the phase-space.

It is well-known that there exists a regular ``bubble of nothing'' or
bubble \cite{Witten:1981gj,Horowitz:2005vp} which carries the same
flux as the black brane. One therefore wonders whether this bubble may
play some role in the phase structure of the black brane. Precisely
based on this consideration, bubble  was considered in the canonical
ensemble in \cite{Lu:2011da} and the phase structure is enriched.
There can be topological transitions among black branes, bubbles and
hot flat space under certain conditions. The bubble really fills
some part of the phase-space though there is still some phase-space
left unoccupied. 

The grand canonical ensemble of black brane system is a little
different. There are no van der Waalsolike phase transitions as in the
canonical ensemble. Since the charge is not fixed in this case,
charged black branes can undergo Hawking--Page-like phase transition
into hot flat space. In \cite{Lu:2010au}, only black branes and hot
flat space are taken into account in the grand canonical ensemble.  
In the present paper, we will discuss the role played by the bubble in
the phase structure.  The bubble  can be
generated from the black brane solution by a double wick rotation in
Minkowski space-time, and in Euclidean space-time the effect is to
interchange one world volume spatial coordinate and the time coordinate
$x\leftrightarrow t$ in the metric, while leaving the form field and
dilaton unchanged. So the bubble geometry is regular and has no
horizon, hence no entropy. One expects that the period of the space
direction $L$ plays a similar role in the bubble case as $\beta$, the
inverse temperature $T$, plays in the black brane system. As a result, the
phase structure depends not only on the temperature and potential
$\Phi$, but also on $L$ as in the canonical ensemble in
\cite{Lu:2011da}. We first follow closely the papers \cite{Lu:2010au}
and \cite{Lu:2011da}.  However, we will find out that in some
conditions there could be phase transition processes where the Gibbs
energy itself is not continuous and the system tends to the boundary
of the cavity. Since our ``zero-loop'' approximation breaks down when
the horizon tends to the boundary in these conditions, this may not be
a right behavior of the system. Nevertheless, if we take these cases
seriously, we will argue that there may be a new phase near the
boundary developed either by quantum effects or other nonperturbative
effects, or the grand canonical ensemble is not stable under certain
circumstances.

In \cite{Lu:2011da}, extremal branes are also considered in the
canonical ensemble and fill some part of the phase-space of $\tilde
d=2,1$ cases.  Extremal branes are very different from the
nonextremal ones, as discussed in
\cite{Gibbons:1994ff,Hawking:1994ii,Teitelboim:1994az}\footnote{\cite{Gibbons:1994ff,Hawking:1994ii,Teitelboim:1994az}
argue that even when the area of the horizon for the extremal black hole is not
zero, the entropy still vanishes. This seems to contradict with the
string theory calculations. Although there are some efferts trying to resolve this
discrepancy, for example \cite{Carroll:2009maa} and the references
therein, the reason for this discrepancy is still not fully understood. However we do not need this result in
our paper since the area of the extremal black brane discussed in this
paper is zero and hence it has no entropy.}. Their thermodynamic properties
may not be obtained simply from the $r_+=r_-$ limit of the nonextremal
ones. For example, they can form equilibria with the environments at
arbitrary temperature. This is because their topology near 
the horizon is different from the nonextremal ones. If we only consider the
Euclidean time direction and the radius direction, the topology of the
nonextremal brane is $R^2$ near the horizon while the one for the
extremal brane is $S^1\times R$. So there is no need to avoid the
conical singularity for the extremal brane and it can have arbitrary
period in the Euclidean time direction hence arbitrary temperature. There is also other evidence
that the extremal black branes can not be seen as the continuous extremal limit
of the nonextremal ones, e.g. \cite{Carroll:2009maa}. The extremal
black branes also
have zero horizon area hence no entropy \cite{Muto:1996xv}. Because of their BPS nature and  zero
entropy property, they may also split into smaller extremal branes with fewer charges without
changing the entropy. However, if we do not care about this, we could
still include the nondilatonic extremal brane in our discussion. We
will see that only when the potential conjugate to the charge at the
boundary is fixed at a certain value can it exist. In this case, it
has zero Gibbs energy which means that it can coexist with the hot
flat space. For dilatonic extremal branes, since the horizon is
already singular, our zero-loop approximation is not
appropriate near the horizon. Therefore we would not consider them in our
discussion.
 
The organization of this paper is as follows: In Sec. 
\ref{sect:potential}, we review the reduced action or the Gibbs free
energy for black brane and obtain the one for the bubble. In Sec. 
\ref{sect:phases}, we discuss the phase structure of this grand
canonical ensemble including black branes, bubbles and the hot flat
space.  The final section contains some discussion on the extremal
cases and the problems we meet.

\section{The action\label{sect:potential}}
Let us recall the black $p$-brane solution in Euclidean signature in
space-time dimension $D=d+\tilde d+2$, $(d=p+1)$
\begin{eqnarray}
ds^{2}&=&\triangle_{+}\triangle_{-}^{-\frac{d}{D-2}}dt^{2}+\triangle_{-}^{\frac{\tilde{d}}{D-2}}(dx^{1})^{2}+\triangle_{-}^{\frac{\tilde{d}}{D-2}}
\sum_{i=2}^p
(dx^{i})^{2}+\triangle_{+}^{-1}\triangle_{-}^{\frac{a^{2}}{2\tilde{d}}-1}d\rho^{2}+\rho^{2}\triangle_{-}^{\frac{a^{2}}{2\tilde{d}}}d\Omega_{\tilde{d}+1}^{2}
\nonumber\\
    A_{[p+1]} &=& - i e^{a\phi_0/2}
\left[\left(\frac{r_-}{r_+}\right)^{\tilde
    d/2} - \left(\frac{r_- r_+}{\rho^2}\right)^{\tilde
    d/2}\right] dt \wedge dx^1\wedge \ldots \wedge dx^p,\nonumber \\
    F_{[p+2]} &\equiv&  dA_{[p+1]}
    = - i e^{a\phi_0/2} \tilde d \,
\frac{(r_- r_+)^{\tilde d / 2}}{\rho^{\tilde d  +
    1}} d\rho \wedge dt \wedge dx^1 \wedge \ldots \wedge dx^p,
\nonumber \\ e^{2(\phi-\phi_0)}&=& \triangle_-^a, 
\label{eq:solution}
\end{eqnarray}
where $a$ is the dilaton coupling and in supergravity theories with
maximal supersymmetry, $a^2=4-\frac{2d\tilde{d}}{D-2}$.
$\triangle_{\pm}$ is defined as $\triangle_{\pm}= 1 -
\left(\frac{r_{\pm}}{\rho}\right)^{\tilde d}$ for $r_+>r_-$, with $r_\pm$
being the two parameters characterizing the solution and related to
the charge and the mass of the black brane.  $\phi_0$ is the
asymptotic value of the dilaton at infinity which is related to the
string coupling $g_s=e^{\phi_0}$. We have extracted one spatial
direction $x^1$ from the sum of the spatial directions. See \cite{Duff:1993ye,Duff:1994an}
for details of the solution.  From the metric, we see that the physical radius of the
$\tilde d+1$ sphere should be $\bar \rho\equiv
\triangle_-^{\frac{a^2}{4\tilde{d}}}\rho$ and  we also define
$\bar{r}_{\pm}\equiv \triangle_-^{\frac{a^2}{4\tilde{d}}}r_{\pm}$. The
charge is calculated to be $Q_d =\frac{\Omega_{\tilde d +1} \tilde
d}{\sqrt{2}\kappa} e^{-a \phi_0/2} (r_+ r_-)^{\tilde d/2}$. 

As mentioned in the introduction, the metric of the bubble is obtained
from the black brane metric just by interchanging the time coordinate
$t$  and one of the world volume spatial coordinate $x^1$
\begin{equation}
ds^{2}=\triangle_{-}^{\frac{\tilde{d}}{D-2}}dt^{2}+\triangle_{+}\triangle_{-}^{-\frac{d}{D-2}}(dx^{1})^{2}+\triangle_{-}^{\frac{\tilde{d}}{D-2}}
\sum_{i=2}^p
(dx^{i})^{2}+\triangle_{+}^{-1}\triangle_{-}^{\frac{a^{2}}{2\tilde{d}}-1}d\rho^{2}+\rho^{2}\triangle_{-}^{\frac{a^{2}}{2\tilde{d}}}d\Omega_{\tilde{d}+1}^{2}.
\end{equation}
The form field $A_{[p+1]}$ and dilaton $\phi$ are kept unchanged. The
space-time is restricted to the $\rho>r_+$ region. The singularity of
the form field at $\rho=0$ is excluded from the geometry and the
charge is the result of the flux around the noncontractible $S^{\tilde
d+1}$.  The period of the $x^1$ direction is chosen to avoid the
conical singularity at $\rho=r_+$ and so the solution is regular.  This
solution can be obtained independently without referring to the black
brane solution by directly solving the classical equation of motion.
It can have the same boundary condition as the black brane when we put
it in to a cavity. Since for an ensemble we only fix the boundary
data, any classical solution that satisfies the boundary conditions
can form an equilibrium with the cavity; therefore one can not exclude
the bubble state from the black brane phase structure.

While
there is an event horizon at $\rho = r_+$ for the black brane, there is no
horizon for bubbles. For black $p$-branes, the inverse of the local
temperature is fixed to be
\begin{equation}
\beta (\bar\rho) = \triangle_+^{1/2}\triangle_-^{-\frac d{2(d+\tilde
d)}}\beta^*=\triangle_+^{1/2}\triangle_-^{-1/\tilde d}\frac{4\pi
\bar r_+}{\tilde d} \left(1 - \frac{\bar r_-^{\tilde d}}{\bar
r_+^{\tilde d}}\right)^{\frac{1}{\tilde d}-\frac 1 2}
\end{equation}
whereas  the local radius of the $x^1$ direction is arbitrary. However, for
bubbles,
the inverse of the local temperature is arbitrary but the local period 
of $x^1$ is
\begin{equation}
L(\bar\rho) =\triangle_+^{1/2}\triangle_-^{-\frac d{2(d+\tilde d)}}
L^*=\triangle_+^{1/2}\triangle_-^{-1/\tilde d}\frac{4\pi \bar
r_+}{\tilde d} \left(1 - \frac{\bar r_-^{\tilde d}}{\bar r_+^{\tilde
d}}\right)^{\frac{1}{\tilde d}-\frac 1 2}
\end{equation}
to avoid the conical singularity. $\beta^*$ and $L^*$ in the above
equations are the inverse temperature and the period seen from
infinity, respectively.

As in \cite{Lu:2010xt,Lu:2010au}, we put the black brane or bubble
inside a cavity with a fixed radius $\bar
\rho=\bar{\rho}_B$. To establish a grand canonical ensemble, we then
fix all the local quantities at the wall of the cavity: the
inverse temperature $\bar \beta$, local period $\bar L$ in $x^1$
direction, local volume $V_{p-1}$ of the $x^i$ $(i=2,\dots, p)$ directions,
dilaton $\phi_B$, and the potential $\bar \Phi$ conjugate to the
charge at the boundary $\bar
\rho_B$. The charge/flux can be now expressed using the boundary data as
$Q_d =\frac{\Omega_{\tilde d +1} \tilde d}{\sqrt{2}\kappa} e^{-a
\phi_B/2} (\bar r_+ \bar r_-)^{\tilde d/2}$ with $\bar r_\pm$
evaluated at the boundary. We can also define the potential $\Phi$ in the
local inertial frame using the form field $A_{[p+1]}\equiv  -i
\sqrt{2} \kappa \Phi d\bar t \wedge d\bar x^1 \ldots d\bar x^p$ where
$(\bar t, \,\, \bar x^1,\,\, \ldots, \bar x^p)$ are the coordinates in
the local inertial frame. So $\Phi$ is the
conjugate potential for $Q$ and one can easily obtain 
\begin{eqnarray}
\bar\Phi=\Phi(\bar \rho_{\rm B}) &=& \frac{1}{\sqrt{2}\kappa} e^{a\phi_{\rm B}/2}
\left(\frac{\bar r_-}{\bar r_+}\right)^{\frac{\tilde d}{2}}
\left(\frac{\triangle_+}{\triangle_-}\right)^{\frac{1}{2}}\Bigg|_{\bar
\rho=\bar\rho_{\rm B}}\,.
\label{eq:Phi}
\end{eqnarray} 
With this setup for the grand canonical ensemble, the classical
Euclidean action for the black brane is obtained in \cite{Lu:2010xt}
\begin{eqnarray}
I_E^{\rm brane}=&-&\frac{\bar\beta \bar
LV_{p-1}\Omega_{\tilde{d}+1}}{2\kappa^2}\bar{\rho}_B^{\tilde{d}}
\left[(\tilde{d}+2)\left(\frac{\triangle_+}{\triangle_-}\right)^{1/2}+\tilde{d}(\triangle_+\triangle_-)^{1/2}-2(\tilde{d}+1)\right]\nonumber
\\
&-&\frac{4\pi
\bar LV_{p-1}\Omega_{\tilde{d}+1}}{2\kappa^2}\bar{r}_+^{\tilde{d}+1}\triangle_-^{-\frac{1}{2}-\frac{1}{\tilde{d}}}\left(1-\frac{\bar{r}_-^{\tilde{d}}}{\bar{r}_+^{\tilde{d}}}\right)^{\frac{1}{2}+\frac{1}{\tilde{d}}}
-\bar \beta \bar LV_{p-1}Q_d\bar \Phi,
\label{eq:IE-brane}
\end{eqnarray}
where we have expressed the volume $V_p=\bar LV_{p-1}$. According to the
zero-loop approximation of the path integral, the Gibbs free energy $G$ can be
obtained from the classical action by
$G=I_E/\bar\beta$ \cite{Gibbons:1976ue}. Since
$G=E-TS-V_p Q_d\bar\Phi$, we can identify the energy $E$ as the first
term in (\ref{eq:IE-brane}) divided by $\bar \beta$, which is
consistent with the ADM mass as $\bar \rho_B\to \infty$ \cite{Lu:1993vt}, and the entropy
$S$ from the second term which is the same as the one obtained in
\cite{Muto:1996xv}.  Since the bubble metric is just obtained by
interchanging $t$ and $x^1$ coordinate, we expect to obtain the bubble
action by interchanging $\bar \beta$ and $\bar L$ in the brane action: 
\begin{eqnarray}I_E^{\rm bubble}=&-&\frac{\bar \beta
\bar LV_{p-1}\Omega_{\tilde{d}+1}}{2\kappa^2}\bar{\rho}_B^{\tilde{d}}
\left[(\tilde{d}+2)\left(\frac{\triangle_+}{\triangle_-}\right)^{1/2}+\tilde{d}(\triangle_+\triangle_-)^{1/2}-2(\tilde{d}+1)\right]
\nonumber \\
&-&\frac{4\pi\bar\beta
V_{p-1}\Omega_{\tilde{d}+1}}{2\kappa^2}\bar{r}_+^{\tilde{d}+1}\triangle_-^{-\frac{1}{2}-\frac{1}{\tilde{d}}}\left(1-\frac{\bar{r}_-^{\tilde{d}}}{\bar{r}_+^{\tilde{d}}}\right)^{\frac{1}{2}+\frac{1}{\tilde{d}}}
-\bar \beta \bar LV_{p-1}Q_d\bar \Phi
\end{eqnarray}
As in the black brane case, one could check that by requiring  the local
minimum of the action with respect to charge $Q_d$ and $r_+$, we
recover the equation of state $\bar L =L(\bar \rho_B)$ and $\bar
\Phi=\Phi(\bar \rho_B)$ for equilibrium. This justifies the validity of this bubble
action. Because of the absence of a horizon, the entropy of the bubble is
zero and hence the sum of the first and the second term divided by
$\bar \beta$ can be identified as the energy of the bubble, which can be
checked to be consistent with the definition of ADM mass as
$\bar\rho_B\to \infty$.

For future convenience, we define the  dimensionless variables
\begin{equation}
x=\left(\frac{\bar{r}_+}{\bar{\rho}_B}\right)^{\tilde{d}},\,\,{b}=\frac{\beta}{4\pi\bar{\rho}_B}\,,\,\,{R}=\frac{L}{4\pi\bar{\rho}_B}\,,\,\,q=\left(\frac{Q_d^*}{\bar{\rho}_B}\right)^{\tilde{d}}\,,\,\,{\varphi}=\sqrt{2}\kappa
e^{-a{\phi_B}/2}\Phi\,.
\end{equation}
where \begin{equation}
Q_d^*\equiv \left(\frac{\sqrt{2}\kappa
Q_d}{\Omega_{\tilde{d}+1}\tilde{d}}e^{a{\phi_B}/2}\right)^{\frac{1}{\tilde{d}}}\,.
\end{equation}
We use the barred variables $\bar b$, $\bar \varphi$, $\bar R$ to denote
the quantities fixed on the boundary. With these dimensionless
variables, one
can rewrite the conditions for equilibrium as 
\begin{equation}\label{eq:b}
\bar{b}=b({x},q),\quad b(x,q)\equiv \frac{1}{\tilde{d}}\frac{x^{1/\tilde{d}}(1-x)^{1/2}}{\left(1-\frac{q^2}{x^2}\right)^{\frac{\tilde{d}-2}{2\tilde{d}}}\left(1-\frac{q^2}{x}\right)^{\frac{1}{\tilde{d}}}}
\end{equation}
for black branes, while for bubbles
\begin{equation}\label{eq:R}
\bar{R}=R({x},q)\,,\quad
R(x,q)\equiv\frac{1}{\tilde{d}}\frac{x^{1/\tilde{d}}(1-x)^{1/2}}{\left(1-\frac{q^2}{x^2}\right)^{\frac{\tilde{d}-2}{2\tilde{d}}}\left(1-\frac{q^2}{x}\right)^{\frac{1}{\tilde{d}}}}\,,
\end{equation}
and 
\begin{equation}
\bar{\varphi}=\varphi({x},q)\,,\quad
\varphi(x,q)\equiv\frac{q}{x}\left(\frac{1-x}{1-\frac{q^2}{x}}\right)^{\frac{1}{2}}\label{eq:phi}
\end{equation}
for both branes and bubbles. 
We first constrain the range of $x$, $q$ to $0\leq q<x<1$,
and hence $0\leq\varphi<1$. In fact, this range can be extend
to $x=q<1$, $\varphi =1$  continuously, which will be considered in
the discussion section. The $x=q=1$ is not well-defined and in this
limit $\varphi$ could have arbitary value. We will argue that this
case can not be physically reached exactly and near this point our
zero-loop approximation is not applicable.

We can also define the reduced action for black branes
\begin{eqnarray}
\tilde I^{\rm brane}_E(\bar b,\bar R,\bar \varphi;q,x) &=& \frac{2\kappa^2 I_E}{(4\pi)^2\bar{
\rho}_B^{\tilde{d} + 2} V_{p - 1} \Omega_{\tilde{d} + 1}}
\nonumber \\&=&-
\bar{b} \bar{R} \left[(\tilde{d} + 2) \left(\frac{1 - x}{1 -
\frac{q^2}{x}}\right)^{1/2} + \tilde{d} (1 - x)^{1/2}\left( 1 -
\frac{q^2}{x}\right)^{1/2} - 2 (\tilde{d} +
1)+\tilde{d}q\bar{\varphi}\right]
\nonumber\\&-&\bar{R}x^{1 + 1/\tilde{d}}
\left(\frac{1 - \frac{q^2}{x^2}}{1 - \frac{q^2}{x}}\right)^{1/2 +
1/\tilde{d}}
\label{eq:I-brane}
\end{eqnarray}
 and by exchanging $\bar{b}$ and $\bar{R}$
we obtain the reduced action for bubbles
\begin{eqnarray}
\tilde I^{\rm bubble}_E(\bar b,\bar R,\bar \varphi;q,x) =&-& \bar{b} \bar{R} \left[(\tilde{d} + 2)
\left(\frac{1 - x}{1 - \frac{q^2}{x}}\right)^{1/2} + \tilde{d} (1 -
x)^{1/2}\left( 1 - \frac{q^2}{x}\right)^{1/2} - 2 (\tilde{d} +
1)+\tilde{d}q\bar{\varphi}\right]
\nonumber \\
&-&\bar{b}x^{1 + 1/\tilde{d}}
\left(\frac{1 - \frac{q^2}{x^2}}{1 - \frac{q^2}{x}}\right)^{1/2 +
1/\tilde{d}}\,.
\label{eq:I-bubble}
\end{eqnarray}
The grand potential or the Gibbs free energy is proportional to the
reduced action divided by $\bar b$. Therefore, for a fixed inverse temperature
$\bar b$, they differ only by a positive constant factor. So, to compare the Gibbs
free energies of the bubble and the black brane at a fixed temperature, 
we can just use the reduced actions instead. Notice that even though we use the
same $x$ in the bubble and brane action, this does not mean that they
are equal. They are independent variables since the physical meanings of $r_+$ for
the black brane and the bubble are different. We also do not differentiate
these two variables in the following discussion and this could be easily understood
from the context.

\section{Phase structure\label{sect:phases}}
If  (\ref{eq:b}), (\ref{eq:R}) and (\ref{eq:phi}) have
solutions for fixed $\bar b$, $\bar R$, $\bar \varphi$, there could be
locally stable black branes or bubbles in the grand canonical
ensemble.  To achieve this, first we can  solve Eq. 
(\ref{eq:phi}) to obtain
\begin{equation}
\frac{q^2}{x^2}=\frac{\bar{\varphi}^2}{1-(1-\bar{\varphi}^2)x}\label{ov}
\end{equation}
and by substituting (\ref{ov}) into $b(x,q)$, $R(x,q)$ in
(\ref{eq:b}) and (\ref{eq:R}), we find
\begin{eqnarray}
b_{\bar{\varphi}}(x)\equiv
b(x,q)=U_{\bar \varphi}(x)\,,&\quad& \text {for black branes,}
\\
\text{and }\quad R_{\bar{\varphi}}(x)\equiv
R(x,q)=U_{\bar \varphi}(x) \,,&\quad&\text { for bubbles,}
\end{eqnarray}
where
\begin{equation}
U_{\bar \varphi}(x)\equiv\frac{x^{\frac{1}{\tilde{d}}}[1-(1-\bar{\varphi}^2)x]^{\frac{1}{2}}}{\tilde{d}(1-\bar{\varphi}^2)^{\frac{1}{2}-\frac{1}{\tilde{d}}}}.
\end{equation}
So to solve (\ref{eq:b}) and (\ref{eq:R}) is to solve  $\bar u=
U_{\bar \varphi}(x)$ for fixed $\bar u$ and $\bar \varphi$, with $\bar
u=\bar b$ for black brane and $\bar u=\bar R$ for bubble, and then
from  (\ref{ov}) we find $q$ for the black brane or the bubble.  Note
that, in general, for $\bar b\neq \bar R$, the solution for the bubble
and the black brane may not be the same. We use $\bar x$ to denote
the solution for the black brane and $\bar y$ for the bubble.
If $\bar x \neq \bar y$, the bubble and the black brane do not have
the same charge/flux from  (\ref{ov}) and one can easily find out that the
one with larger $x$ solution carries more charge/flux. 

\begin{figure}
\centering
\subfigure[]{\label{fig:unstable}
\includegraphics[width=7cm]{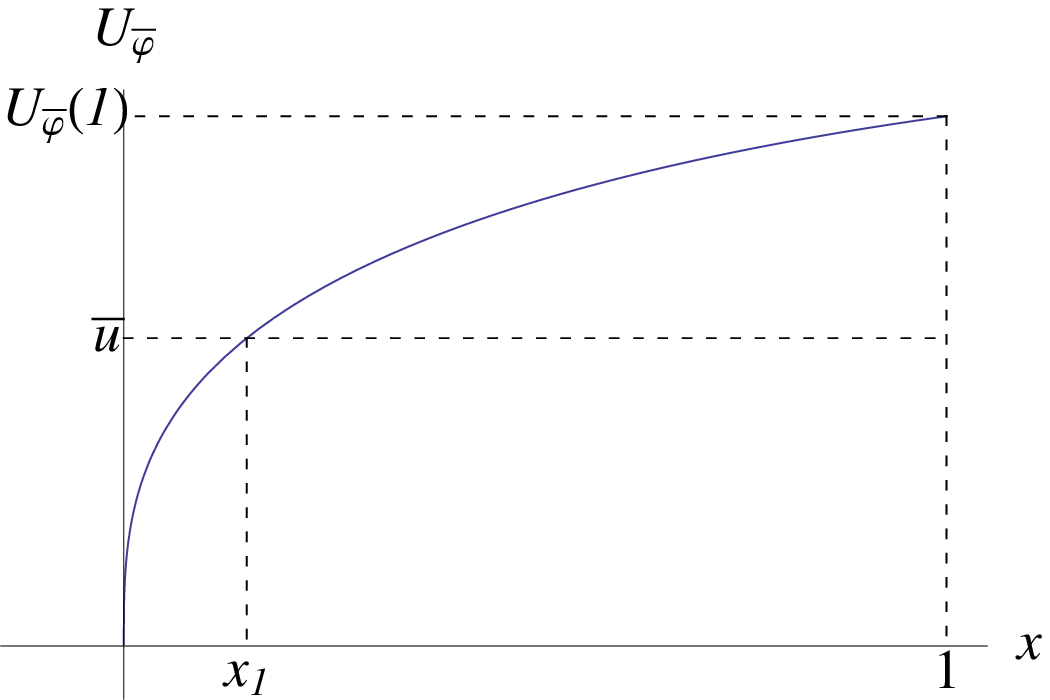}
}
\subfigure[]{\label{fig:localstable}\includegraphics[width=7cm]{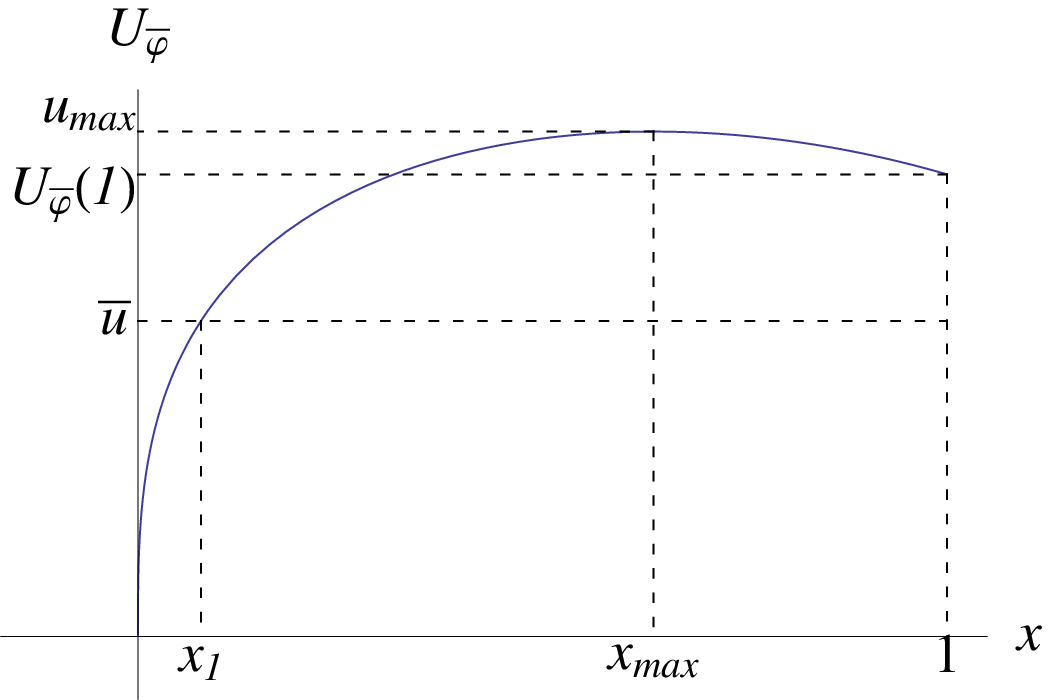}}
\subfigure[]{\label{fig:globalstable}\includegraphics[width=7cm]{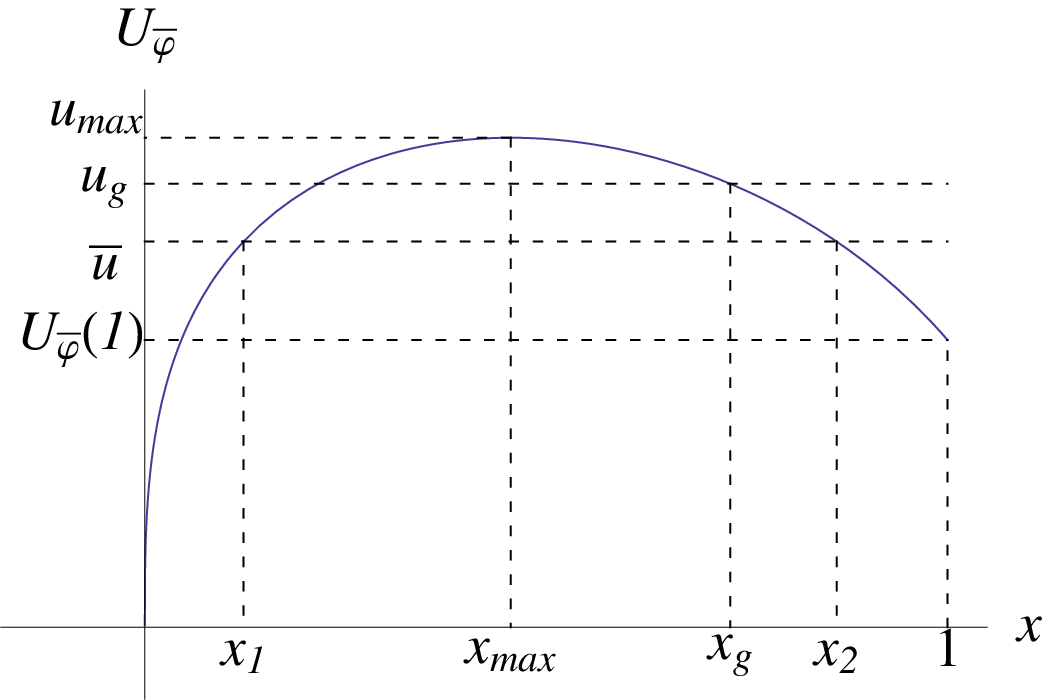}}
\caption{The $U_{\bar \varphi}$ vs $x$ diagrams for different $\bar
\varphi$: (a) for $\sqrt{\frac {\tilde d}{2+\tilde d}}<\bar
\varphi \leq 1$. (b) for $\frac {\tilde d}{2+\tilde d}<\bar
\varphi< \sqrt {\frac {\tilde d}{2+\tilde d}}$.
(c) for $ \bar
\varphi< \frac {\tilde d}{2+\tilde d}$.}
\end{figure}

The discussion on the solutions of $\bar u=
U_{\bar \varphi}(x)$ 
under different conditions has already been done in \cite{Lu:2010au} for
the black brane, and can be directly used in the bubble case.  The
results can be summarized as follows,
\begin{enumerate}
\item For $\sqrt {\frac {\tilde d}{2+\tilde d}}<\bar
\varphi < 1$, $U_{\bar \varphi}(x)$ is monotonically increasing for
$0<x<1$. If $\bar u>U_{\bar\varphi} (1)$ there is no solution, and
if $0<\bar u< U_{\bar \varphi} (1)$, there is one unstable solution.
The $U_{\bar\varphi}$ vs $x$ graph is shown in Fig. \ref{fig:unstable}.
\label{unstable1}
\item For $\bar
\varphi< \sqrt {\frac {\tilde d}{2+\tilde d}}$, there is a local
maximum for $U_{\bar\varphi} (x)$ with
\begin{eqnarray}
u_{\rm max}& =& \left(\frac{2}{2+\tilde
d}\right)^{\frac{1}{\tilde d}} \left[\tilde d (\tilde d +2) \left(1
- \bar
    \varphi^2\right)\right]^{-\frac{1}{2}}
\nonumber \\&<&
\frac{1}{\sqrt{2\tilde d}}\left(\frac{2}{2+\tilde
    d}\right)^{\frac{1}{\tilde d}}
\label{eq:umax}
\end{eqnarray}
at 
\begin{eqnarray} x_{\rm
max} = \frac{2}{(2 + \tilde d)(1 - \bar\varphi^2)}.
\label{eq:xmax}
\end{eqnarray} 
\begin{enumerate}
\item When $\frac {\tilde d}{2+\tilde d}<\bar
\varphi< \sqrt {\frac {\tilde d}{2+\tilde d}}$, 
the $U_{\bar\varphi}$ vs $x$ graph is shown in Fig. \ref{fig:localstable}.
\begin{enumerate}
\item $0<\bar u<U_{\bar \varphi}(1)$, there is one unstable solution.
\label{unstable2}
\item  $U_{\bar \varphi}(1)<\bar u <
u_{\rm max}$, there are two solutions: the
smaller is unstable, and the larger is locally stable with $\tilde I_E>0$.
\label{localStable1}
\end{enumerate}
\item For $\bar
\varphi< \frac {\tilde d}{2+\tilde d}$, we define $\bar x_g = \frac{4(\tilde d +1)}{(\tilde d
+2)^2 \left(1-\bar
    \varphi^2\right)} $ where $u_g=U_{\bar \varphi}(\bar x_g)=\frac{\left(4(\tilde d
+1)\right)^{\frac{1}{\tilde d}}}{(\tilde d +2)^{1
    + \frac{2}{\tilde d}} \sqrt{1- \bar \varphi^2}}$ and $\tilde
I(\bar x_g)=0$.
The $U_{\bar\varphi}$ vs $x$ graph for this case is shown in Fig. \ref{fig:globalstable}.
\begin{enumerate}
\item $0<\bar u<U_{\bar \varphi}(1)$, there is one unstable solution
with $\tilde I_E>0$.
\label{unstable3}
\item For $u_g<\bar u < u_{\rm max}$, there are two solutions: 
the smaller is unstable, and the larger is locally stable with $\tilde
I_E>0$.\label{localStable2}
\item For $U_\varphi(1)<\bar u <u_g $, there are two solutions:
the smaller is unstable, and  
the larger is locally stable with $\tilde
I_E<0$.\label{globalStable}
\end{enumerate}
\end{enumerate}
\end{enumerate}
By changing $\bar u$ into $ \bar b$ or $\bar R$, we can obtain the
property of the black brane or the bubble respectively. Whether the
final state of the system is a black brane, a bubble or the hot flat
space depends on which one has the smallest reduced action. There
are four different cases.  The first one, if  both $\bar b$ or $\bar
R$ are chosen in cases \ref{unstable1}, \ref{unstable2},
\ref{localStable1}, \ref{unstable3}, \ref{localStable2}, both black
brane and bubble are either unstable or locally stable with $\tilde
I_E>0$. Since the hot flat space is an equilibrium state with
$\tilde I_E=0$, and we suppose there is no other unknown equilibrium
with $\tilde I_E<0$ in these cases, the system will tend to the hot
flat space, in line with the results of \cite{Lu:2010au}. However, there is a
problem in this claim, because the hot flat space may not be the
global minimum of the Gibbs energy. We will come to this point in the
discussion section.
The second one, if $\bar b$  is chosen in cases \ref{unstable3} or
\ref{localStable2} but $ \bar R$  is chosen in case
\ref{globalStable}, the locally stable bubble will have $\tilde I_E<0$ and the black
brane is either unstable or has $\tilde I_E>0$. Thus, the final state
will be the bubble. There could also be a problem in this claim which
will be discussed later. The third case is interchanging $\bar b$ and
$\bar R$, bubble and black brane in the previous case. The final
complicated case is when both $\bar b$ and $\bar R$ are chosen in case
\ref{globalStable} where both bubble and black brane have negative
reduced actions.  In this case we have to compare these two negative
ones.

\begin{figure}
\includegraphics{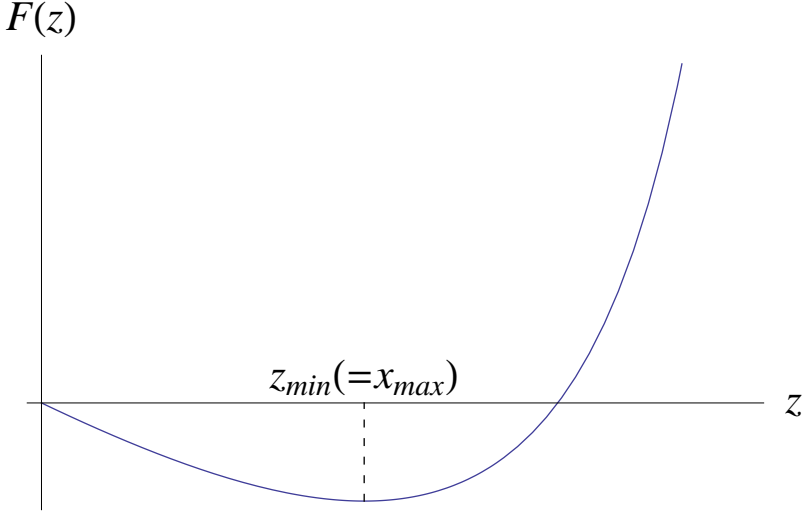}
\caption{\label{fig:F1} The $F(z)$ is increasing for $z>z_{\rm
min}=x_{\rm max}$. }
\end{figure}

For this purpose, we can express the reduced on-shell action for the black
brane and the bubble in the same form: for the black brane
\begin{equation}
\tilde I^{\rm brane}_E=-\bar{b}\bar{R}F_{\bar{\varphi}}(\bar{x})
\end{equation}
while for the bubble
\begin{equation}
\tilde I^{\rm bubble}_E=-\bar{b}\bar{R}F_{\bar{\varphi}}(\bar{y})
\end{equation}
where
\begin{equation}
F_{\bar{\varphi}}(z)=(\tilde{d}+2)\sqrt{[1-(1-\bar{\varphi}^2)z]}+\frac{\tilde{d}}{\sqrt{[1-(1-\bar{\varphi}^2)z]}}-2(\tilde{d}+1)
\end{equation}
with $0\leq\bar\varphi<\tilde d/(\tilde d +2)$ and $0<z<1$, $\bar x$
and $\bar y$ being the solutions for  (\ref{eq:b}) and (\ref{eq:R})
for the black brane and  the bubble respectively. $F_{\bar{\varphi}}(z)$ has a minimum
at $z_{\rm min}=\frac{2}{(2+\tilde d)(1-\bar\varphi^2)}$, which is
equal to $x_{\rm max}$ in (\ref{eq:xmax}), and is increasing in the
interval $z_{\rm min}<z<1$ while decreasing in $0<z<z_{\rm min}$ as
indicated in Fig. \ref{fig:F1}.  As a result, of the bubble and the black
brane, the larger one will have smaller action. Since the monotonic
property of $b$ vs $x$
or $R$ vs $x$ is depicted in the same diagram in
Fig. \ref{fig:globalstable} for fixed $\bar \varphi$, which is
decreasing for $x>x_{\rm max}=z_{\rm min}$, the black brane will be
larger, i.e. $\bar x>\bar y$, when $\bar R > \bar b$, and inversely,
the bubble will be larger when $\bar b>\bar R$.  Therefore, when $\bar
R > \bar b$ the black brane will have smaller reduced action and when
$ \bar R < \bar b$, the reduced action for the bubble is smaller.

Finally, the possible phase structure of this grand canonical ensemble can
be summarized as follows,
\begin{enumerate}
\item When $\frac{\tilde d}{\tilde d +2}<\bar \varphi$, the final
state is the hot flat space.\label{flat1}
\item When $0<\bar \varphi<\frac{\tilde d}{\tilde d +2}$, 
\begin{enumerate}
\item When $\bar b\in (0,U_{\bar \varphi}(1))\cup(u_g,+\infty)$ and
$\bar R \in(0,U_{\bar \varphi}(1))\cup(u_g,+\infty)$, the final state
is the hot flat space.\label{flat2}
\item When $\bar b\in(U_{\bar \varphi}(1),u_g)$ and $\bar R\in
(0,U_{\bar \varphi}(1))\cup(u_g,+\infty)$ , the final state is the
black brane. Reciprocally, when $\bar b\in (0,U_{\bar
\varphi}(1))\cup(u_g,+\infty)$ and $\bar R\in(U_{\bar
\varphi}(1),u_g)$, the final state is the bubble.
\item When $U_{\bar \varphi}(1)<\bar b<\bar R<u_g$ the final state is
the black brane. Inversely, when $U_{\bar \varphi}(1)<\bar R<\bar b
<u_g$, the final state is the bubble.
\item There could be some coexisting states:
\begin{itemize}
\item Three components coexisting point: $\bar b=\bar R=u_g$.
\item The bubble and black brane coexisting phase: $U_{\bar \varphi}(1)<\bar b=\bar
R<u_g$.
\item The black brane and hot flat space coexisting phase: $\bar
b=u_g$, for $\bar R>u_g$ or $\bar R\in (0,U_{\bar \varphi}(1))$. And the
bubble and hot flat space coexisting phase: $\bar R=u_g$,  for
$\bar b>u_g$ or $\bar b\in (0,U_{\bar \varphi}(1))$.
\end{itemize}
\end{enumerate}
\end{enumerate}

\section{Discussion\label{sect:discuss}}
In the previous section we have discussed the possible phase structure of the
grand canonical ensemble for the bubble, black brane, hot flat
space system. Compared with the system with only black brane and
hot flat space, the phase structure depends not only on the
potential $\bar \Phi$ (or the dimensionless $\bar \varphi$),
temperature $T$ (or inverse temperature $\bar\beta$, dimensionless
$\bar b$) but also on the size of one world volume spatial direction
$\bar R$. 
However
there could be some region of the phase-space where our zero-loop
approximation fails and we must exclude this region from our
discussion. To see this, let us recount the phase structure here in another way. If we fix
$\tilde d/(\tilde d+2)<\bar\varphi$, there is nothing interesting but
the hot flat space.  If we fix $0<\bar \varphi<\tilde d/(\tilde
d+2)$, there could be different phase transition processes:
\begin{enumerate}
\item If we fix the temperature such that $\bar b\in (0,U_{\bar
\varphi}(1))\cup(u_g,+\infty)$ and increase the size of the $x^1$ 
 direction from small to large, at first the system will be
the chargeless hot flat space. Then just above $\bar R= U_{\bar \varphi}(1)$ there
is a phase transition to  bubble with $\bar r_+$ close to the boundary of
the cavity and the charge close to the extreme value. Put
another way, the bubble suddenly appears from the boundary of the
cavity. This is an instantly charging process with the Gibbs free energy
jumping discontinuously from zero to a negative one, which characterizes a
``zeroth-order'' phase transition.  Then as $\bar R$ increases, $\bar
r_+$ and the charge are decreasing gradually while the Gibbs energy is
increasing. After $\bar R$ increases to $\bar R = u_g$, where $\bar r_+$
and  $Q_d$ decrease to the corresponding values at $x_g$ and the Gibbs
energy increases to zero, the system begins to discharge  and
transforms to the hot flat space through a coexisting phase of the
bubble and the ``hot flat space''.  This is a first order transition
with continuous Gibbs energy. Then the system will stay in the ``hot
flat space''. 
\item If we fix the temperature such that $\bar b\in (U_{\bar
\varphi}(1),u_g)$ and do the same thing, at first the system is a black
brane, of which the horizon size and the charge do not change as $\bar R$
increases. There will be a phase transition to bubble  at just above $\bar
R=U_{\bar \varphi}(1)$ where the charge suddenly increases to near
extremal one and the $\bar r_+$ also suddenly rises to near boundary. In
other words, the same as in the previous case, the near extremal bubble
appears from the boundary.  Similar to the process in the previous case,
the Gibbs energy decreases discontinuously which is a
``zeroth-order'' phase transition. Then, as we increase $\bar R$, the
bubble becomes smaller and discharges gradually. When $\bar R$ is
raised to $\bar R =\bar b$, where the size of the bubble has shrunk
to the same size as the original black brane, the bubble begins to
transform back to the black brane through their coexisting phase. This
is also a first order transition. After
that, the system remains in the same black brane phase, i.e. the
charge and the horizon size no longer change.
\end{enumerate}
We can also fix $\bar R$ and raise $\bar b$ as in above two processes,
and the consequence is that the black brane and the bubble will
exchange their roles.

We have seen from above two cases that there are ``zeroth-order'' phase
transition processes where the Gibbs free energies are not continuous.
In fact, in these cases, for $\bar R<U_{\bar \varphi}(1)$, the Gibbs
free energy around $x=1$ for the bubble, although not being at the
stationary points, have already dropped below zero or
below the one for the original black brane. Similarly, in the cases
when we fix $\bar\varphi\in (\tilde d/(\tilde d+2),1)$ where only the
``hot flat space'' exists, as the
temperature is raised to a certain value,
the reduced action near $x=1$, still not
being at the stationary points, will also drop below zero. It seems
that, if the Gibbs energy here is correct for this thermodynamic
system in these situations, the ``hot flat space'' is not a global
minimum of the Gibbs free energy, and the system tends  to $x=1$
bubbles or black branes which are not the stationary points of the Gibbs
free energy at first sight. Nevertheless, from (\ref{ov}), there seems
to be an extremal solution for $x=q=1$ with arbitrary $\bar\varphi<1$.
Since extremal branes can form equilibria with the environments having
arbitrary temperature and $\bar R$
\cite{Gibbons:1994ff,Hawking:1994ii,Teitelboim:1994az}, one may think
that this extremal brane (from the extremal limit of
(\ref{eq:solution}), there is no difference between the bubble and
the brane)  will be the final steady state in all these cases.  However,
this is a spurious solution of (\ref{ov}).  To see this, we define
$r=(\bar \rho/\bar \rho_B)^{\tilde d}$ and look at $\varphi$ at
arbitrary $\bar \rho$ which is just  (\ref{eq:Phi}) with $\bar
\rho_{\rm B}$ changed to $\bar \rho$, and using  (\ref{ov}), we find:
\begin{equation}
\varphi(r,x)=\frac {\bar\varphi(r-x)^{1/2}}{(r(1-x)+x\bar
\varphi^2(r-1))^{1/2}}\,.
\label{eq:phi-r}
\end{equation}
From this equation, when $\bar r_+=\bar \rho_{\rm B}$ which is the
extremal case with $x=1$, the only solution has $ \varphi(r\to 1,
1)=1$ and at the equilibrium, $\bar\varphi=1$, which means that the
$x=q=1$ extremal one can only have $\bar\varphi=1$ boundary condition
in equilibrium. This contradicts with the previous arbitary $\bar
\varphi$ solution.
In consequence, the solution with
$x=q=1$ and arbitrary $\bar \varphi$ could not be physically realized
exactly.  If we first set $r=1$ exactly (which is $\bar \rho = \bar \rho_{\rm
B}$) and $x<1$, i.e.  we set
the boundary condition first and then choose $\bar r_+<\bar \rho_{\rm
B}$, we find the boundary condition $\varphi(1,x)=\bar \varphi$ which
could be arbitrary.
  This implies that the $x=q=1$ solution with arbitrary $\bar
\varphi$ can only be seen as a limiting case which can not be reached
exactly. This argument is similar to the one used in
\cite{Zaslavsky:1996zz}. So we should exclude this phase from the phase structure.
Let us look at what happens near this extremal case but not reaching
it exactly.
For black branes, as $x$ and $q$ are approaching $1$, the curvature singularity
at $r_-$ is coming closer to the horizon and also closer to the wall
of the cavity.
Since the quantum effect will be essential near the singularity, our
zero-loop approximation is not applicable in these
situations. The Gibbs free energy near $x=1$ may be modified by
quantum corrections. Similarly, for bubbles, when $x,q\to1$ quantum
effects must be important near $\rho=r_+$ and the wall. So, these
zeroth-order phase transitions may not be the correct behavior of
the system and are just indications of the failure of our method in
describing the system under these circumstances.  One possibility of
the system near $x=1$ is that there is a new stationary point near
$x=1$ developed by quantum effects or by some other unknown
topological solutions, which may lift the Gibbs free energy near
$x=1$. If the new stationary point is a global minimum, there will be
a new phase.  The other possibility is that there are no new stationary
points and the system itself is not
well-defined or is unstable, which means we can not set up a steady
grand canonical ensemble under such circumstances.  

It is worth mentioning that  some solutions with both
bubbles and black holes present were found in five dimensinal pure
gravity \cite{Elvang:2002br,Elvang:2004iz} and can be uplifted to string
theory by embedding in 11 dimensions  and a series of dualities
\cite{Harmark:2004ws,VazquezPoritz:2011zr}. There could be
configurations of branes connected by KK bubbles. It will be
interesting to investigate whether the solutions obtained this way
could play a role in the phase-space of the black brane and may also
be related to the issue raised above. We will leave this possibility
as a future research direction.

After excluding the $x=q=1$ extremal case, we can also study the
extremal nondilatonic brane with $x=q<1$ whose geometry at the horizon is not
singular, for example, $M2$ brane and $M5$ branes in M-theory, and $D3$
brane in string theory. From 
(\ref{eq:I-brane}), we see that it has zero entropy automatically
 because of its zero horizon area which is already known in \cite{Muto:1996xv}.
From (\ref{ov}) we obtain $\bar \varphi=1$ and hence the
Gibbs energy is zero for any $\bar b$ from 
(\ref{eq:I-brane}). Thus, it has the same Gibbs free energy as the hot
flat space. Therefore, only at $\bar \varphi=1$ could the extremal
brane exist and it can also coexist with the hot flat space. We
should point out that this is a zero-loop result. If the higher-order
 fluctuations are taken into account, the Gibbs energy of the
extremal brane and the hot flat space may be different and these
two states may not coexist. Since in our paper, we are considering
only the leading-order approximation, we will list the coexisting
state as a possible phase. As is mentioned in the introduction section
that 
the extremal brane can not be seen as a continuous extremal limit of the
nonextremal one, it must appear as a result of some noncontinuous
processes such as pair production or quantum tunneling. But the
dynamics behind these processes is not clear to us and is beyond the
scope of this paper. As for
the extremal dilatonic brane, since the space-time geometry at the horizon is
singular, our method does not apply here and hence we exclude them from the
phase structure. 

So, if we take into account the discussion in this section, the phase
structure listed at the end of the previous section would be modified.
First, another extremal brane and hot flat space coexisting phase
with $\bar \varphi=1$ should be added to the phase structure.
Second, in some regions of the phase-space, the final states of the
system is not clear to us due to the limitation of the method and we
should exclude these regions from our discussion. In
particular, when $\bar \varphi>\frac {\tilde d}{\tilde d+2}$ and
either $\bar b$ or $\bar R$ is less than some value about which the
zero-loop approximation is not applicable, we are not sure what
happens. We could estimate this value to be $ (1-\bar
\varphi^2)^{1/2+1/\tilde d}/(2(\tilde d +1)(1-\bar \varphi))$ by
requiring the reduced action to be zero at $x=1$.  For $0<\bar
\varphi<\tilde d/(\tilde d+2)$, we are still not sure about the phases
when either $\bar b$ or $\bar R$ is in $(0,U_{\bar \varphi}(1))$. We
can not say too much about the final phases in these parts of phase-space at
the present stage. This problem could be left for future study. 

\section{Acknowledgement}
We thank J.~X.~Lu for helpful discussion. Z. X. is  
supported by the Fundamental Research Funds for the Central
Universities under grant No. WK2030040020. C. Wu and J. Xu are supported
by grants from the Chinese Academy of Sciences, a grant from 973
Program with grant No. 2007CB815401 and a grant from the NSF of China
with Grant No. 10975129. 

\end{document}